# Impenetrable barriers for positrons in neighbourhood of superheavy nuclei with Z>118

**V P Neznamov**[1]
Federal state unitary enterprise Russian Federal Nuclear Center – All-Russian Research Institute of Experimental Physics (FSUE "RFNC-VNIIEF"), 37 Mira pr., Sarov, Nizhny Novgorod region, 607188, Russia

E-mail: vpneznamov@vniief.ru

**Abstract.** Analysis of quantum mechanical motion of charged half-spin particles in the repulsive Coulomb field by means of second-order equation results in that an impenetrable potential barrier not explored earlier was found. For a particle at rest with a reduced mass $m$, the barrier radius is equal to half classical radius: the barrier radius decreases with increase in the particle energy. For the stable and quasi-stable nuclei with $Z > 118$, presence of an impenetrable barrier as $\beta^+$-decay leads to the existence of "traps" for positrons in the neighbourhood of nuclei and as $Z_{cr} \simeq 170$ (with emission of electron-positron pairs by vacuum) leads to the existence of a quasi-constant source of annihilation quanta.

## 1. The equations for fermions in the Coulomb field

*1.1. The Dirac equations in the Coulomb field*
Here and below, we will generally use the system of units $\hbar = c = 1$.
Bispinor wave function for stationary states.

$$\Psi(\mathbf{r},t) = \begin{pmatrix} \varphi(\mathbf{r}) \\ \chi(\mathbf{r}) \end{pmatrix} e^{-iEt}. \qquad (1)$$

$E$ is the energy of the Dirac particle.
In spherical coordinates $(r, \theta, \varphi)$

$$\varphi(\mathbf{r}) = e^{im_\varphi \varphi} \xi(\theta) F(r),$$
$$\chi(\mathbf{r}) = e^{im_\varphi \varphi} (-i) \sigma^3 \xi(\theta) G(r). \qquad (2)$$

In (2) $F(r), G(r)$ are the radial wave functions; $m_\varphi$ is the azimuthal quantum number; $\sigma^k$ are the Pauli matrices; the spinor $\xi(\theta) = \begin{pmatrix} {}_{-1/2}Y(\theta) \\ {}_{+1/2}Y(\theta) \end{pmatrix}$, where ${}_{-1/2}Y(\theta), {}_{+1/2}Y(\theta)$ are spherical harmonics for spin ½.

---

[1] vpneznamov@vniief.ru

The dimensionless variables: $\rho = r/l_c$; $\varepsilon = E/m$; the Coulomb field $V(\rho) = \pm Z\alpha/\rho$; $l_c = 1/m$ is the Compton wavelength of the Dirac particle; $\alpha = e^2/\hbar c$ is the constant of the fine structure; $Z$ is the number of protons in atomic nucleus.

The separation constant $\kappa = \mp 1, \mp 2 ... = \begin{cases} -(l+1), & j = l+1/2 \\ l, & j = l-1/2 \end{cases}$; $j, l$ are the quantum numbers of the angular and orbital momenta of the Dirac particle.

The Dirac equations for components of bispinor $\Psi(\mathbf{r})$

$$(\varepsilon + 1 - V(\rho))\chi(\rho,\theta,\varphi) = \boldsymbol{\sigma}\mathbf{p}\varphi(\rho,\theta,\varphi),$$
$$-(\varepsilon - 1 - V(\rho))\chi(\rho,\theta,\varphi) = \boldsymbol{\sigma}\mathbf{p}\chi(\rho,\theta,\varphi). \tag{3}$$

After separation of variables: The Dirac equations for radial wave functions are

$$(\varepsilon + 1 - V)G(\rho) = \frac{dF}{d\rho} + \frac{1+\kappa}{\rho}F,$$
$$-(\varepsilon - 1 - V)F(\rho) = \frac{dG}{d\rho} + \frac{1-\kappa}{\rho}G, \tag{4}$$

*1.2. Second-order equation for spinor $\varphi(\mathbf{r})$*

$$\left((\varepsilon - V(\rho))^2 - \mathbf{p}^2 - ie\boldsymbol{\alpha}\mathbf{E}\right)\Psi(\rho,\theta,\varphi) = 0 \text{ [1]},$$
$$\alpha^k = \begin{pmatrix} 0 & \sigma^k \\ \sigma^k & 0 \end{pmatrix}; \quad e\mathbf{E} = -\nabla V, \tag{5}$$

From equations (3)

$$\chi = \frac{1}{\varepsilon + 1 - V}\boldsymbol{\sigma}\mathbf{p}\varphi, \tag{6}$$

and

$$\left((\varepsilon - V)^2 - \mathbf{p}^2 - \frac{1}{\varepsilon + 1 - V}ie(\boldsymbol{\sigma}\mathbf{E})(\boldsymbol{\sigma}\mathbf{p})\right)\varphi(\rho,\theta,\varphi) = 0. \tag{7}$$

Self-adjoint second-order equation can be represented in the form

$$\frac{1}{\rho^2}\frac{d}{d\rho}\left(\rho^2\frac{d}{d\rho}\right)\psi(\rho) + \left((\varepsilon - V)^2 - 1 - \frac{\kappa(\kappa+1)}{\rho^2} - \frac{3}{4}\frac{\left(\frac{dV}{d\rho}\right)^2}{(\varepsilon + 1 - V)^2} - \frac{1}{2}\frac{\frac{d^2V}{d\rho^2}}{(\varepsilon + 1 - V)} + \frac{1}{(\varepsilon + 1 - V)}\frac{1}{\rho}\frac{dV}{d\rho}\kappa\right)\psi(\rho) = 0. \tag{8}$$

For self-conjugacy in (7), a non-unitary transformation is implemented [2]

$$\psi(\rho) = \frac{\rho}{\sqrt{\varepsilon + 1 - V}}F(\rho) \tag{9}$$

In case of the repulsive Coulomb field $V(\rho) = Z\alpha/\rho$, there is a value $\rho = \rho_{cl}$ at which $(\varepsilon + 1 - V(\rho_{cl})) = 0$. At that

$$\rho_{cl} = Z\alpha/(\varepsilon + 1) \tag{10}$$

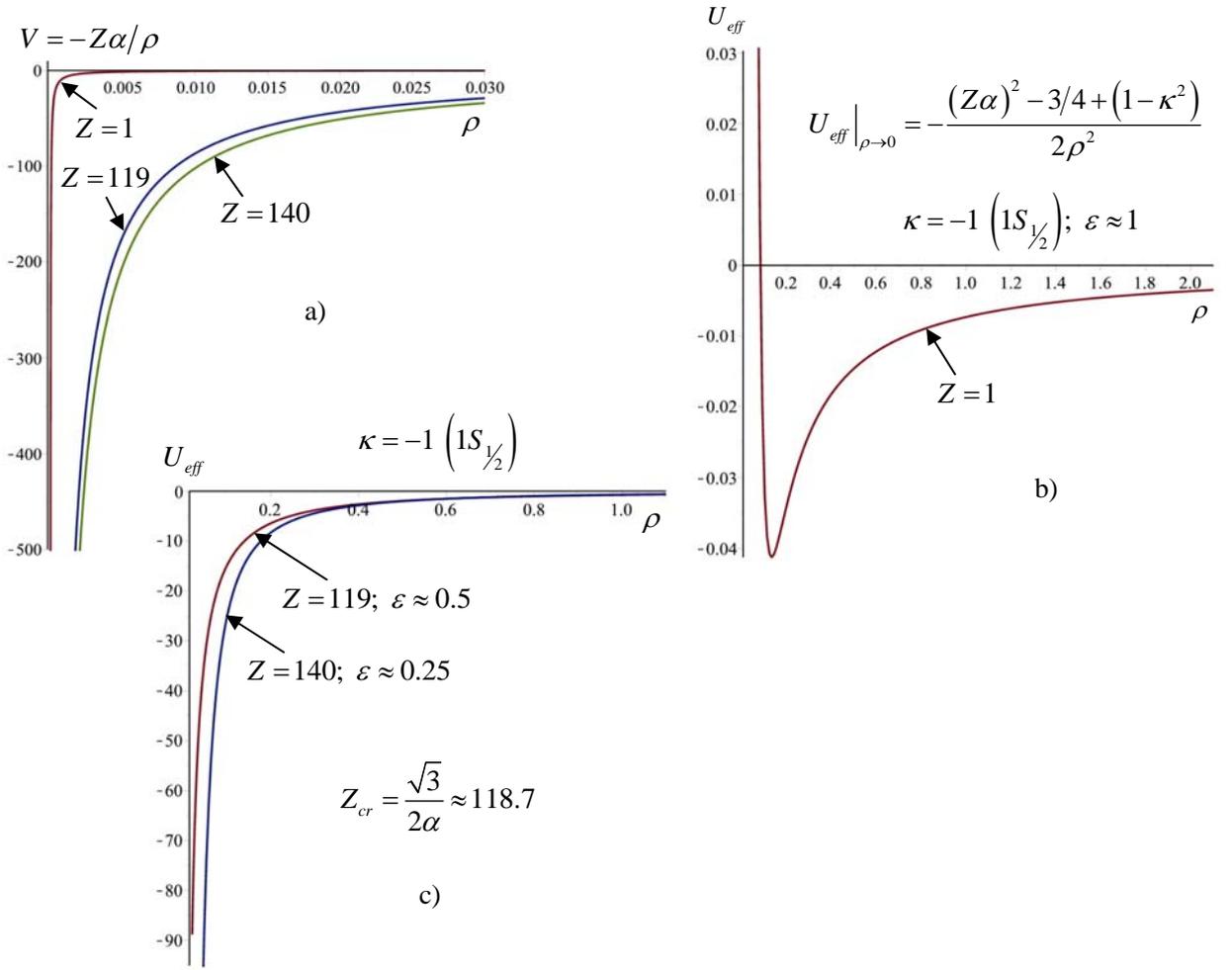

**Figure 1.** a) the attracting Coulomb field $(V = -Z\alpha/\rho)$; the effective potentials of second-order equation (8): b) as $Z < Z_{cr}$, c) as $Z > Z_{cr}$.

In dimensional units

$$r_{cl} = Z\frac{e^2}{mc^2}\bigg/\left(1+\frac{E}{mc^2}\right). \tag{11}$$

For $\psi(\rho) = f(\rho)/\rho$ equation (8) is

$$\frac{d^2 f}{d\rho^2} + 2\big(E_{Schr} - U_{eff}(\rho)\big)f = 0, \tag{12}$$

where

$$E_{Schr} = \frac{1}{2}\big(\varepsilon^2 - 1\big) \tag{13}$$

and

$$U_{eff}\big|_{\rho\to\rho_{cl}} = \frac{3}{8}\frac{1}{(\rho-\rho_{cl})^2}. \tag{14}$$

Barrier (14) is quantum mechanically impenetrable. If $f\big|_{\rho\to\rho_{cl}} = (\rho-\rho_{cl})^s \sum_k f_k (\rho-\rho_{cl})^k$ then from the inditial equation for (12) $s_1 = 3/2;\ s_2 = -1/2$. Solution $s_2 = -1/2$ is unphysical because of

square-nonintegrability of the wave function $f(\rho)$. Then $f|_{\rho \to \rho_{cl}} = f_0 (\rho - \rho_{cl})^{3/2}$ and in accordance with (9) $F|_{\rho \to \rho_{cl}} = C(\rho - \rho_{cl})^2$; $F|_{\rho \to \rho_{cl}}$, $\dfrac{dF}{d\rho}\bigg|_{\rho \to \rho_{cl}} = 0$; $G|_{\rho \to \rho_{cl}} = C_1$, $\dfrac{dG}{d\rho}\bigg|_{\rho \to \rho_{cl}} = C_2$.

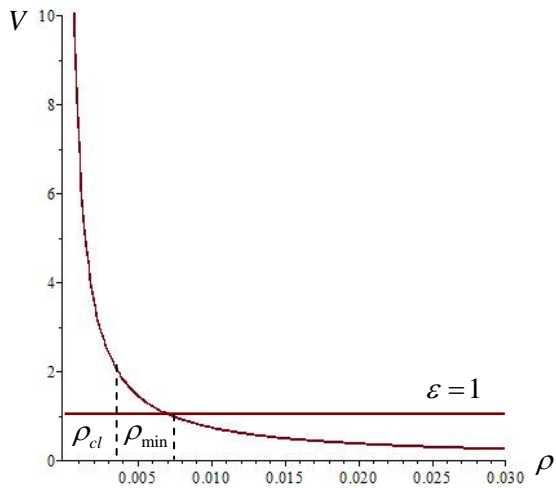

a) $V(\rho) = Z\alpha/\rho$ at $Z=1$ (for $\varepsilon = 1$, $\rho_{min} = 1/137$, $\rho_{cl}^e = 1/274$)

$F(\rho_{cl}) = 0$; $\dfrac{dF}{d\rho}\bigg|_{\rho = \rho_{cl}} = 0$

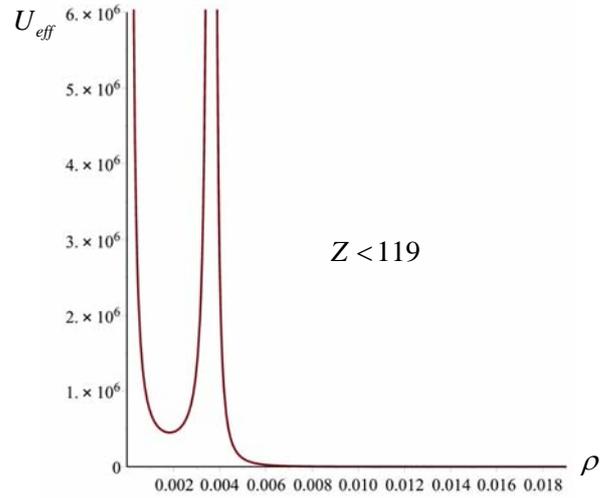

b) $Z<Z_{cr}$ ($Z=1$, $\kappa = -1$, $\varepsilon = 1$, $\rho_{cl} = 0.00365$)

$f(\rho_{cl}) = 0$

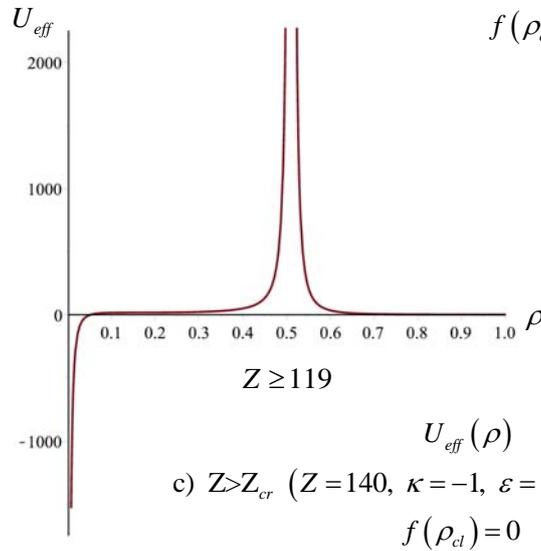

c) $Z>Z_{cr}$ ($Z=140$, $\kappa = -1$, $\varepsilon = 1$, $\rho_{cl} = 0.51$)

$f(\rho_{cl}) = 0$

**Figure 2.** The curves $V(\rho)$ and $U_{eff}(\rho)$ in case of the repulsive Coulomb field.

Consequences of existence of impenetrable barrier
1. The existence of an impenetrable barrier in an electron motion in the repulsive Coulomb field, in terms of the scattering theory, means the existence of a well-defined impact parameter $p$

$$p = r_{cl}. \qquad (15)$$

When probing the internal structure of an electron in all possible experiments with electron-electron scattering, the impossibility to penetrate towards radii $r < r_{cl}$ should be recorded.

In this case, the value $r_{cl}$ decreases with the growth of the electron energy:

$$r_{cl} \approx \frac{Zr_e}{E} m_e c^2 \quad \text{at } E \gg m_e c^2. \tag{16}$$

In (16), $r_e$ is classical radius of electron $\left(r_e = e^2/mc^2\right)$.

Let us emphasize that this conclusion holds true only within the applicability of one-particle quantum mechanics.

2. For motion of positrons, the impenetrable barrier (14) exists in the case of the Coulomb field of the atomic nuclei of the Periodic system. For fully ionized stable atoms, the $Ze$ variation range is from $Z = 1$ (Hydrogen ion) to $Z = 92$ ($U^{238}$ nucleus). For a positron at rest in the center-of-mass system, the barrier radius varies from $1.4 \cdot 10^{-13}$ cm $(Z = 1)$ to $1.3 \cdot 10^{-11}$ cm $(Z = 92)$. Such change can be establish experimentally that would confirm the existence of discussed barrier and the applicability of one-particle quantum mechanics for this problem. At subsequent hypothetical increase $Z$ up to $Z = Z_{scr} = (170 - 177)$, an energy level of electron ground state reaches the upper limit of the lower continuum $E = -m$. Then, in case of electron-free $S$-states on $K$-shell, a spontaneous production of two electron-positron pairs are occurred [3].

The electrons occupy the free $K$-shell, the atomic charge decreases by two units $((Z_{scr} - 2)e)$, positrons leave the atomic system where, in principal, they can be recorded. However, taking into account an impenetrable barrier (14), the positrons will be in potential well within the barrier with $r = r_{cl}$ (see Fig. 2). Hereafter the positrons annihilate with electrons of the $K$-shell with emission of quanta and the system reverts to the initial state $(Z = Z_{scr})$ with further event recurrence. For the outer world, the source of the annihilation of electrons with positrons will be the sole information source for the atomic system with $Z = Z_{scr}$.

3. The existence of the impenetrable barrier indirectly revealed the relationship between the spin and the charge of particles.
   Indeed:
   - for spinless particles, there is no barrier;
   - when signs of charges the spin particle and the Coulomb potential are opposite, there is no barrier;
   - when the spin particle and the Coulomb potential have like charges, the barrier is present, and arrangement of the barrier is changed at change of particle energy.

Future models of the substructure should quantitatively describe specified above the relationship between the spin and the charge of particles.

## 2. Second-order equation and Foldy-Wouthuysen representation

For the spinor $u(\mathbf{r})$, the equation has the form of

$$\left[ (E - eA_0)^2 - (\mathbf{p} - e\mathbf{A})^2 - m^2 + e\boldsymbol{\sigma}\mathbf{H} - \frac{1}{E - eA_0 + m} ie\boldsymbol{\sigma}\mathbf{E}\boldsymbol{\sigma}(\mathbf{p} - e\mathbf{A}) \right] u(\mathbf{r}) = 0. \tag{17}$$

In case of stationary states, the electromagnetic potentials $A_0(\mathbf{r}), A^k(\mathbf{r})$ are independent of time. Let $A_0(\mathbf{r}) = 0$, $A^k(\mathbf{r}) \neq 0$. Then equation (17) is self-adjoint and

$$Eu(\mathbf{r}) = \left( \pm \sqrt{m^2 + (\mathbf{p} - e\mathbf{A})^2 + e\boldsymbol{\sigma}\mathbf{H}} \right) u(\mathbf{r}). \tag{18}$$

Then, let us consider the case of $A^k(\mathbf{r}) = 0, A_0(\mathbf{r}) \neq 0$. In this case, the latter addendum in (17) is a nonself-adjoint operator. Let us perform the non-unitary transformation of the similarity of the equation (17) and spinor $u(\mathbf{r})$

$$\Phi(\mathbf{r}) = g u(\mathbf{r}), \tag{19}$$

where

$$g = (E - eA_0 + m)^{-1/2}. \tag{20}$$

As a result, equation (17) is reduced to the form:

$$g\left[(E - eA_0)^2 - \mathbf{p}^2 - m^2 - \frac{1}{(E - eA_0 + m)} i\boldsymbol{\sigma}\mathbf{E}\boldsymbol{\sigma}\mathbf{p}\right] g^{-1} \Phi(\mathbf{r}) = 0 \tag{21}$$

and, finally, to

$$\left[(E - eA_0)^2 - \mathbf{p}^2 - m^2 - \frac{3}{4} \frac{1}{(E - eA_0 + m)} \mathbf{E}^2 + \frac{1}{2} \frac{1}{E - eA_0 + m} \mathrm{div}\mathbf{E} + \frac{1}{E - eA_0 + m} \boldsymbol{\sigma}(\mathbf{E} \times \mathbf{p})\right] \Phi(\mathbf{r}) = 0. \tag{22}$$

Equation (18) with $A^k(\mathbf{r}) \neq 0, A_0(\mathbf{r}) = 0$, written actually in the Hamiltonian form and can be easily compared with the Dirac equation in the Foldy-Wouthuysen representation (FW) [4].
In case of $A^k(\mathbf{r}) = 0, A_0(\mathbf{r}) \neq 0$, it is impossible to write equations (17), (22) in the closed Hamiltonian form. It can be done only by the method of successive approximations.

*2.1. Foldy-Wouthuysen chiral representation*
If we present bispinor as

$$\psi(\mathbf{r}, t) = \begin{pmatrix} \psi_L(\mathbf{r}) \\ \psi_R(\mathbf{r}) \end{pmatrix} e^{-iEt}, \tag{23}$$

then the addendum $\beta m$ in the Dirac equation mixes spinors $\psi_L(\mathbf{r}), \psi_R(\mathbf{r})$ and the Dirac equation with the non-zero mass of fermions does not possess chiral symmetry.
Conversely, in second-order equation there are no addendums mixing spinors $\psi_L, \psi_R$ and it can be written as

$$\left[(p_0 - eA_0)^2 - (\mathbf{p} - e\mathbf{A}_0)^2 - m^2 + \boldsymbol{\sigma}\mathbf{H} - ie\boldsymbol{\sigma}\mathbf{E}\right]\psi_L = 0, \tag{24}$$

$$\left[(p_0 - eA_0)^2 - (\mathbf{p} - e\mathbf{A}_0)^2 - m^2 + \boldsymbol{\sigma}\mathbf{H} + ie\boldsymbol{\sigma}\mathbf{E}\right]\psi_R = 0. \tag{25}$$

Equations (24), (25) are chiral-symmetric. Presence or absence of fermion mass will have no effect on chiral symmetry (24), (25). Noteworthy that equations (24), (25) allow obtaining a closed expression for the Hamiltonian FW in chiral representation

$$H_{FW} = eA_0 + \gamma_5 \sqrt{m^2 + (\mathbf{p} - e\mathbf{A})^2 + e\boldsymbol{\sigma}\mathbf{H} - ie\boldsymbol{\alpha}\mathbf{E}}. \tag{26}$$